\documentclass[a4paper,12pt,english]{article}

\usepackage{amsfonts,bm,amssymb,euscript,array,babel,cite}
\usepackage{mathtools}
\usepackage{tikz-cd}
\usepackage{amsmath}
\usepackage{tensor}
\usepackage{hhline}
\usepackage[amsmath]{empheq}
\usepackage{hyperref}
\usepackage{cancel}
\usepackage{mathrsfs}
\usepackage{pdfsync}
\usepackage{multirow}
\usepackage{tikz}

\usepackage{framed}

\newsavebox{\mysaveboxM}
\newsavebox{\mysaveboxT}

\makeatletter

\newcommand{\be}{\begin{equation}}
\newcommand{\ee}{\end{equation}}

\def \bea{\begin{eqnarray}} 
\def\eea{\end{eqnarray}}

\def\bi{\begin{itemize}} 
\def\ei{\end{itemize}}

\makeatother

\def\one{\mbox{1 \kern-.59em {\rm l}}}

\numberwithin{equation}{section}

\begin{document}

\makeatother
\parindent=0cm
\renewcommand{\title}[1]{\vspace{10mm}\noindent{\Large{\bf #1}}\vspace{8mm}} \newcommand{\authors}[1]{\noindent{\large #1}\vspace{5mm}} \newcommand{\address}[1]{{\itshape #1\vspace{2mm}}}

\begin{titlepage}

\begin{flushright}
	RBI-ThPhys-2022-43
\end{flushright}

\begin{center}

\title{ {\large {Instances of Higher Geometry in Field Theory}}}

  \authors{\large Athanasios {Chatzistavrakidis}{\footnote{Athanasios.Chatzistavrakidis@irb.hr}} 
  }
 
 
  \address{ 
  	 Division of Theoretical Physics, Rudjer Bo\v skovi\'c Institute \\ Bijeni\v cka 54, 10000 Zagreb, Croatia \\

 }

\vskip 2cm

\begin{abstract}
Generalisations of geometry have emerged in various forms in the study of field theory and quantization. This mini-review focuses on the role of higher geometry in three selected physical applications. After motivating and describing some basic aspects of algebroid structures on bundles and (differential graded) Q-manifolds, we briefly discuss their relation to ($\alpha$) the Batalin-Vilkovisky quantization of topological sigma models, ($\beta$) higher gauge theories and generalized global symmetries and ($\gamma$) tensor gauge theories, where the universality of their form and properties in terms of graded geometry is highlighted. 

\end{abstract}

\end{center}

\vskip 2cm

\end{titlepage}

\setcounter{footnote}{0}

\newpage

\section{Introduction}\label{sec1}

The interaction between physics and geometry has a long and fertile history and has played a central role in the understanding and the precise description of physical phenomena. Fundamentally, all physical phenomena refer either to the gravitational interaction, described by General Relativity (GR) and its limits, or to the interactions in particle and condensed matter physics, collectively described through Quantum Field Theory (QFT). Both GR and QFT have a geometrical character, through Riemannian geometry and the geometry on (principal) fiber bundles, but also strong ties with the concept of symmetry, 
which is intimately related to the study of groups and algebras. 

Geometry and symmetry find a unified description via higher (or generalised) geometry.{\footnote{In this mini-review, the word ``higher'' is preferred. One reason is that it is sometimes used here as an umbrella term for a variety of different---but often closely related---ideas, including generalised complex geometry, differential geometry of homotopy algebras or algebroids, graded supergeometry and noncommutative geometry, although focus will be on a small subset due to length restrictions. }} 
Higher geometry introduces a series of unifying frameworks, which often sprang with motivation from specific physical problems. Poisson and symplectic geometry, the backbones of classical mechanics and the springboards for quantization, were the early driving forces. The study of constrained Hamiltonian systems {\footnote{It is useful to recall that all gauge field theories are constrained Hamiltonian systems.}} and integrability conditions led to the notion of Dirac structure  which clarifies the relation between Poisson and presymplectic structures \cite{cour} and paved the road to the development of Courant algebroids \cite{Liu:1995lsa} and, via complexification, to generalised complex geometry \cite{Hitchin:2003cxu,Gualtieri:2003dx}, which brings together symplectic and complex geometry as well as interpolating structures therein. From a mathematical standpoint, these are instances of unifying the notions of algebra (corresponding to symmetry) and vector bundle (corresponding to geometry) to the notion of an \emph{algebroid}, originally introduced in the 60s \cite{MR0216409}.
Remarkably, from a physical standpoint, these structures have made their appearance in a variety of ways in QFT, notably in the context of quantization via the Batalin-Viskovisky formalism, in describing higher gauge theories that exhibit higher (global and/or gauge) symmetries and/or dualities, but also in string, M and string field theories.  

The purpose of this mini-review, part of a volume with title Noncommutativity in Physics, is to collect together developments regarding generalizations of geometry and their applications in physics which although distinct from noncommutative geometry and field theory they are closely related to it and often refer to similar physical problems. (The common origins of generalized geometry and noncommutativity were explored in \cite{Jurco:2013upa}.) Here, motivated by certain problems in constrained Hamiltonian systems, field theory and quantization, we focus on the higher geometry of algebroids and differential graded manifolds (Q-manifolds) and discuss their emergence in physics through three selected applications: ($\alpha$) topological sigma models and their Batalin-Vilkovisky (BV) quantization, ($\beta$) higher gauge theories and their relation to  generalized global symmetries, and ($\gamma$) gauge theories for general tensor fields, including spins $> 1$. 
Obviously there are more things omitted than included in our discussion, for some of which the reader can consult the rest of the papers in this volume.

\section{Roads to Higher Geometry}
\label{sec2}

There are two main routes one can follow in describing higher geometries, namely ($\alpha$) ordinary differential geometry on general or extended vector bundles and ($\beta$) graded geometry on (super)manifolds. The two approaches are complementary, with different merits and contextual usefulness. In the following we briefly describe them starting from their physical motivations.

In classical mechanics and in the Hamiltonian formulation, one encounters generalised coordinates and momenta which together form the phase space of a mechanical system and  obey Hamilton's equations. In a coordinate-independent way, one introduces a Poisson bracket of functions on the phase space. Given two functions $F$ and $G$, their Poisson bracket $\{F,G\}$ is another function and the bracket obeys the antisymmetry condition and the Jacobi identity. This can be described using an antisymmetric 2-vector $\Pi\in \Gamma(\wedge^2TM)$ on a smooth manifold $M$ that satisfies 
\begin{equation}
	[\Pi,\Pi]_{\text{SN}}=0\,, \label{Poisson}
\end{equation}
in terms of the Schouten-Nijenhuis bracket of multivector fields, which receives a $p$-vector and a $q$-vector and gives a $(p+q-1)$-vector as result. In the case of 2-vectors, the result is an antisymmetric 3-vector and specifically Eq. \eqref{Poisson} in a coordinate basis results in 
\begin{equation}
	\Pi^{l[i}\partial_l\Pi^{jk]}=0\,,
\end{equation}
where antisymmetrization is taken with weight 1. This is nothing but the Jacobi identity of the Poisson bracket. Then $\Pi$ is called a Poisson 2-vector and $(M,\Pi)$ a Poisson manifold.

Modelling the phase space of a Hamiltonian system as a Poisson manifold, endows it with a natural symplectic structure. One should be cautious, however, in that when constraints are present in the system, the Poisson structure can be singular and the symplectic structure degenerate, in which case it is referred as presymplectic. A presymplectic manifold is then one equipped with a closed 2-form $\omega$, 
\begin{equation}
	\mathrm{d} \omega=0\,,
\end{equation}
not necessarily nondegenerate. One question then is, what is the relation between Poisson and presymplectic structures and are there interpolations  between them? The answer was given in \cite{cour}, using an extension of the Lie bracket of vector fields known as the Courant bracket, given as 
\begin{equation}
	[X_1+\eta_1,X_2+\eta_2]=[X_1,X_2]+{\cal L}_{X_1}\eta_2-{\cal L}_{X_2}\eta_1-\frac 12 \mathrm{d}\left(X_1(\eta_2)-X_2(\eta_1)\right)\,,
\end{equation}
where $X_1$ and $X_2$ are vector fields and $\eta_1$ and $\eta_2$ are 1-forms. This is then a skew-symmetric binary operation locally on the extension of the tangent bundle of $M$ by its cotangent bundle, $E=TM\oplus T^{\ast}M$. The properties of this bracket were organised in an axiomatic form originally in \cite{Liu:1995lsa} and given the name of a Courant algebroid, which is a quadruple $(E,[\cdot,\cdot],\langle\cdot,\cdot\rangle,\rho:E\to TM)$ of a vector bundle over $M$, a (non Lie, but in the original definition skew-symmetric) bracket, a nondegenerate symmetric bilinear form and a smooth bundle map called ``anchor'', respectively.{\footnote{Other versions corresponding to the same notion exist, for example based on Leibniz-type brackets (a.k.a. Dorfman bracket) \cite{DR,Severa:2017oew} or foliations \cite{Chen_2013}.}}

Returning to the discussion on constrained Hamiltonian systems, Poisson and presymplectic structures are both associated with subbundles of $E$ where the Courant bracket is closed and the bilinear form vanishes. These are called Dirac structures and they are specific instances of Lie algebroids. The latter are mergers of vector bundles and Lie algebras in the sense that they correspond to vector bundles $L$ with a Lie algebra structure on their sections, given by a Lie bracket. They comprise $L$, the bracket $[\cdot,\cdot]_{L}$ on its space of sections and an anchor $\rho:L\to TM$ which is a homomorphism of bundles and participates in the Leibniz rule 
\begin{equation}
	[e_1,fe_2]_{L}=f[e_1,e_2]_{L}+\rho(e_1)f\, e_2\,.
\end{equation}
Apart from Dirac structures, simple---and in a sense extremal---examples of Lie algebroids include (i) Lie algebras, which are Lie algebroids over a point $M=\{\ast\}$, and (ii) the tangent bundle Lie algebroid, where $E=TM$ itself, the anchor is the identity map and the bracket is the ordinary one for vector fields, which are the sections in this case. We note that this is not the only possible Lie algebroid structure over $TM$, just the most common one. Apart from these examples, it is worth mentioning another simple instance, the action Lie algebroid. This is modelled on the trivial vector bundle $M\times \mathfrak{g}$, where $\mathfrak{g}$ is a Lie algebra acting on $M$. It is a case repeatedly encountered in physics in the context of gauge theory, when there is an action of a symmetry algebra. We will encounter some more complicated examples below.  

It is useful to note that since the Courant bracket is an extension of the ordinary Lie bracket of vector fields, which is symmetric under diffeomorphisms, this symmetry is inherited by the former and it is augmented by an additional symmetry that corresponds to gauge symmetry for a 2-form \cite{Gualtieri:2003dx}. This is the primary reason why this structure has naturally appeared in theories that contain a Riemannian metric and a Kalb-Ramond field, notably in  string theory. What is more, the fiber metric that corresponds to the inner product $\langle\cdot,\cdot\rangle$ of the Courant algebroid is of split signature and introduces an $O(d,d;\mathbb{R})$ structure, $d=\text{dim}\,M$. This is the reason why Courant algebroids have played a central role in modern developments regarding T-duality and string theory backgrounds with generalized fluxes, often in the context of double field theory. For all these developments, we refer to several useful reviews that highlight different aspects of the topic and contain further pointers to the literature \cite{Plauschinn:2018wbo,Aldazabal:2013sca,Hohm:2013bwa,Berman:2013eva}. 

The second route refers to graded geometry. Loosely this means that we consider spaces whose local coordinates carry some additional degree. More mathematically precise treatments may be found in \cite{Qiu:2011qr,CATTANEO_2011}---here we follow a more physics oriented approach, according to the requirements of the present review. The idea of graded coordinates is certainly not unfamiliar in theoretical physics. Supersymmetry and supersymmetric field theories are the prime examples of this, especially in the superspace formulation. There we encounter bosonic (degree-0) and fermionic (degree-1) coordinates, thus having a $\mathbb{Z}_2$ grading. 

Another, more general context where such grading is encountered is the Batalin-Vilkovisky (BV) formalism. This is the most general method to organise the local symmetry structure and subsequently quantize a theory with gauge symmetries. It generalises the familiar Fadeev-Popov method of handling redundancies in non-Abelian gauge theory and it is necessary in very general cases where the gauge theory features one or more of the following three properties: ($\alpha$) a gauge algebra that closes only on-shell (meaning after the classical field equations of the theory are taken into account)---this is often referred as an open gauge algebra, ($\beta$) a gauge algebra whose structure constants are field-dependent---this is sometimes called a soft algebra, ($\gamma$) a reducible gauge symmetry, one where not all gauge parameters are independent, typically arising when differential forms of degree $\ge$ 2 are present, as for example in higher gauge theory. According to the BV formalism, to handle such cases one must introduce a series of ghosts, ghosts for ghosts, antifields and antighosts and also additional trivial pairs to the classical basis. All these objects carry an additional degree to their parity, which can be any integer, thus introducing a finer $\mathbb{Z}$ grading. The field space then is a graded supermanifold with coordinates of different degrees. Detailed expositions of the formalism are found for example in \cite{HT,Gomis:1994he,Barnich:2000zw}. 

The basic idea in graded geometry is that instead of tensor fields (sections on vector bundles) one can shift attention to functions on graded supermanifolds. The prototypical example is the degree-shifted tangent bundle of a manifold $M$, denoted $T[1]M$ and indicating that the fibers are assigned shifted degree 1. The local coordinates on such a graded manifold are $x^{\mu}$ and $\theta^{\mu}$ of degrees 0 and 1 respectively. Then $\theta^{\mu}$ are anticommuting 
\begin{equation}
	\theta^{\mu}\theta^{\nu}=-\theta^{\nu}\theta^{\mu}\,.
\end{equation} In a certain sense, the coordinates of degree 1 have replaced the differentials $\mathrm{d}x^{\mu}$ and the product of functions the wedge product of differential forms. In other words, a differential form can be alternatively described as a function on $T[1]M$ instead of as a section of the exterior algebra of $M$. 

What is then the relation of the above picture to higher geometry? The most straightforward way to answer this is roughly that algebroid structures are in one-to-one correspondence with suitable graded manifolds equipped with an odd (degree 1 and homological) vector field $Q$ that satisfies $Q^2:=\frac 12 \{Q,Q\}=0$, where the curly brackets are the anticommutator,---also called $Q$-manifolds or differential graded manifolds. To be more precise, a Lie algebroid given by the triplet of structures $(L,[\cdot,\cdot]_{L},\rho)$ can be as well described as the Q-manifold $(L[1],Q_L)$ of the degree-shifted vector bundle $L$ and a homological vector field given as 
\begin{equation}\label{QLA}
	Q_L=\rho^{\mu}{}_a(x)\xi^{a}\frac{\partial}{\partial x^{\mu}} -\frac 12 C^{a}{}_{bc}(x)\xi^{b}\xi^{c}\frac{\partial}{\partial\xi^{c}}\,,
\end{equation}
where $x^{\mu}$ are degree 0 coordinates on $M$ and $\xi^{a}$ the degree 1 coordinates on the fiber of $L$. The nilpotence of this vector field is then identical to the defining conditions of a Lie algebroid once the function $\rho^{\mu}{}_{a}(x)$ is identified with the components of the anchor map in some basis and the functions $C^{a}{}_{bc}(x)$ with the structure functions of the Lie bracket in the same basis. 

In a similar fashion, a Courant algebroid given by the data $(E,[\cdot,\cdot],\langle\cdot,\cdot\rangle,\rho)$ can be described as a Q-manifold $(T^{\ast}[2]T[1]M,Q)$, which is a generalised phase space with graded generalized coordinates and momenta of degrees 0, 1, 1 and 2. In fact, such a graded manifold is more than just a Q-manifold. Being a phase space, it carries a natural odd symplectic structure (referred as a P-structure) that is moreover compatible with the Q-structure in the sense that the flow of the corresponding odd symplectic form $\omega$ is invariant along the homological vector field, ${\cal L}_{Q}\omega=0$. Such differential graded manifolds are called graded symplectic supermanifolds \cite{DR} or symplectic $L_2$ (read $L_{n}$ with $n=2$) algebroids \cite{sometitle} or QP-manifolds \cite{Alexandrov:1995kv}. The precise relation between Courant algebroids and $L_{\infty}$ (strong homotopy Lie) algebras is found in \cite{Roytenberg:1998vn}---see also \cite{Grewcoe:2020ren} for an alternative approach. 

QP-manifolds are of course not a structure specific to Courant algebroids. First of all, they already exist at the level of Lie algebroids. The prototypical example is the degree-shifted cotangent bundle $T^{\ast}[1]M$ with coordinates $x^{\mu}$ and $\xi_{\mu}$, with $\xi$ being essentially the conjugate momenta of $x$. Equipping it with a suitable homological vector field with structure functions $\Pi^{\mu\nu}(x)$ as $\rho$ and $\partial_{\rho}\Pi^{\mu\nu}$ as $C$, nilpotence is equivalent to the 2-vector being a Poisson structure and the correspondence is to a Lie algebroid on the cotangent bundle with anchor induced by this Poisson structure and Lie bracket given by the Koszul-Schouten bracket of 1-forms. It is useful to point out here that we have described Poisson geometry in four alternative ways up to this point, 
\begin{equation}
	(M,\{\cdot,\cdot\}) \quad \text{or} \quad (M,\Pi) \quad \text{or} \quad (T^{\ast}M,[\cdot,\cdot]_{\text{KS}},\Pi^{\sharp}) \quad \text{or} \quad (T^{\ast}[1]M,Q)\,,     
\end{equation} 
corresponding to the Poisson bracket, Poisson 2-vector, cotangent Lie algebroid and Q-manifold respectively. 
QP-structures exist on other underlying graded manifolds too, for example on $T^{\ast}[n]T[1]M$ for $n\in \mathbb{Z}$, $T^{\ast}[n]T^{\ast}[1]M$, $T^{\ast}[n](\wedge^{n-1}[n-1])T[1]M$ and so on, see e.g. \cite{Ikeda:2012pv,Bouwknegt:2011vn}.   

Although QP-manifolds are important on their own right due to their direct relation to the BV formalism and topological field theory \cite{Alexandrov:1995kv}, as will be discussed below, the relation of algebroids and Q-manifolds is more fundamental. This can be seen in a variety of ways, which we briefly highlight through examples. There exists a Lie algebroid over the cotangent bundle with the anchor being a \emph{twisted} Poisson structure in the sense of \cite{Severa:2001qm} and the bracket being the twisted Koszul-Schouten one. The twist in this case corresponds to a closed, not necessarily exact, 3-form on $M$. The corresponding graded manifold has a Q but not a QP-structure. The same conclusion holds for 4-form-twisted Courant algebroids (also called pre-Courant algebroids{\footnote{This and some relaxed structures thereof were used in \cite{Chatzistavrakidis:2018ztm} for membrane sigma models.}}) \cite{Hansen:2009zd,Bruce_2019,Vaisman:2004msa,Liu_2016} but also in more general cases, for example for $(p+2)$-form twisted R-Poisson manifolds \cite{Chatzistavrakidis:2021nom}. Even more explicitly, an arbitrary Lie algebroid modelled on $E[1]$ does not have a P-structure at all, let alone a QP one.

\section{Selected Applications} 
\label{sec3}

\subsection{Topological Sigma Models \&  BV formalism}\label{sec31}

Topological sigma models were introduced in \cite{Witten:1988xj} and their early history is summarized in \cite{Birmingham:1991ty}. From a general perspective, they are topological quantum field theories described as maps from a source manifold, the world volume, to a target manifold. In 2D, there exists a variety of topological sigma models, with prominent cases Witten's A and B topological string models. The quantization of such models follows the BV formalism. One may then ask what is the geometry behind this very general quantization scheme. Notably, it turns out that for topological sigma models, a solution to the classical  master equation corresponds to a QP structure on the  space of fields \cite{Alexandrov:1995kv,Schwarz:1992nx,Schwarz:1992gs,Witten:1991zz}.  

This correspondence between the higher geometry of QP manifolds and topological quantum field theory goes by the name of the AKSZ construction. Remarkably, this method turns a complicated problem into a simple geometrical one. Indeed, determining the solution to the classical master equation can be very demanding, especially as one increases the dimension of the target or the degree of reducibility of the gauge algebra. In the AKSZ construction, once the graded source is chosen, typically being the graded tangent bundle $T[1]\Sigma$ of the world volume $\Sigma$, and the target space is identified as a QP manifold, the solution is straightforward. The classical action functional of the topological sigma model is replaced by one where all fields are promoted to superfields, functions on $T[1]\Sigma$. These superfields are the pull-backs of the graded coordinates of the QP manifold by an extended sigma model map and their components contain all ghosts and antifields of the gauge theory, suitably organized.  

Alluding to the examples of QP manifolds mentioned above, considering $T^{\ast}[1]M$ one obtains the Poisson sigma model, originally introduced in \cite{Schaller:1994es,Ikeda:1993fh}, whose direct relation to the A-model, also with respect to their observables, is described in \cite{Bonechi:2007ar,Bonechi:2016wqz}. This means that this field theory encodes Poisson geometry in its gauge structure. It is worth mentioning that the quantum BV action of the Poisson sigma model on a disk was used in \cite{Cattaneo:1999fm} to provide a physical derivation of the Kontsevich solution to the problem of deformation quantization on Poisson manifolds \cite{Kontsevich:1997vb}, which served as the stepping stone for the development of several physical models based on noncommutative geometry, the central topic of this volume. It should also be mentioned that the A-model also serves as the starting point of the proposed quantization via branes \cite{Gukov:2008ve}, based on its A-branes  \cite{Kapustin:2001ij}, which has been suggested as a systematic method to address the problem of quantizing a symplectic manifold.  

Going back to the examples of QP manifolds, the next one is the graded second order bundle $T^{\ast}[2]T[1]M$, which is related to Courant algebroids as already discussed. From the topological quantum field theory viewpoint, this target space corresponds to Chern-Simons theory in 3D, or more precisely an extension thereof called Courant sigma model \cite{Ikeda:2002wh,Hofman:2002jz,Roytenberg:2006qz}, which can be viewed as a coupled Chern-Simons-BF theory. In other words, the consistency of the gauge structure of this field theory is equivalent to the local coordinate expressions for the axioms of a Courant algebroid. Aside these two basic models in 2D and 3D, one can construct further analogs in higher dimensions too \cite{sometitle}. 

However, it is interesting to emphasize that not all topological sigma models come together with a QP structure on the target space that can be directly pulled back to the space of fields. Notable examples of topological field theories  where this is not the case are models with Wess-Zumino term, such as the completely gauged G/G Wess-Zumino-Witten model or the 3-form-twisted Poisson sigma model \cite{Klimcik:2001vg} based on the twisted Poisson structure of \cite{Severa:2001qm} and its corresponding Lie algebroid mentioned earlier. The latter model, even though it does have a Q and a P structure, the QP structure is obstructed since the symplectic form is not Q-invariant in general. Moreover, this is not an isolated 2D example. It has a semi-infinite class of higher-dimensional analogs, given by the twisted R-Poisson sigma models in $p+1$ dimensions with a closed $(p+2)$-form as Wess-Zumino term \cite{Chatzistavrakidis:2021nom}. In a different direction, there also exists a class of 2D topological sigma models that interpolate between the G/G WZW and the twisted Poisson, called Dirac sigma models \cite{Kotov:2004wz} and having no P-structure in general. 

One may then ask, what is the solution of the classical master equation for such models and how about the geometry of the BV formalism in such cases? Although there is no general answer to these questions yet, a few recent developments exist. Regarding the twisted Poisson sigma model in 2D, the solution to the classical master equation was found in \cite{Ikeda:2019czt}, where it is also shown that a naive extension of the AKSZ method does not give the right answer.  Nonetheless, the solution involves a number of definite higher geometric ingredients, notably a Lie algebroid $E$-connection with nonvanishing $E$-torsion and an associated basic $E$-curvature. Lie algebroid $E$-connections and their associated $E$-covariant derivatives generalize their ordinary counterparts such that the arguments are sections of $E$ instead of vector fields. Specifically an $E$-covariant derivative on $E$ is a map $^{E}\nabla_e: E\to E$ such that 
\begin{equation}
	^{E}\nabla_e(fe')= f\,^{E}\nabla_ee'+ \rho(e)f\, e'\,,
\end{equation}
where $e,e'$ are sections of $E$ and $\rho$ the anchor.{\footnote{We note that $E$-covariant derivatives on a different vector bundle $V$ can also be defined.}} Using this notion, one may define the associated torsion and curvature tensors as in usual differential geometry, see e.g. \cite{Kotov:2016lpx}.
This higher $E$-geometry is instrumental in establishing the global structure of the topological sigma model beyond local coordinate patches. In a similar fashion, the solution to the classical master equation has been recently found for topological Dirac sigma models \cite{Chatzistavrakidis:2022wdd} in 2D and for 4-form-twisted R-Poisson sigma models in 3D \cite{Chatzistavrakidis:2022hlu}. In the former case, the higher $E$-geometry has been determined and it involves two separate $E$-connections with their corresponding higher torsion and curvature tensors. In the latter case, the higher geometry has not been fully developed yet.     

\subsection{Higher Form Gauge Theories \&  Global Symmetries}
\label{sec32}

Higher gauge theory refers to QFTs that contain in their field content differential forms of degree higher than 1, as opposed to ordinary gauge theory where fields are 1-forms (``vectors''), connections on a principal bundle. As already mentioned, having higher form fields in your gauge theory leads to reducible gauge symmetry and therefore the BV formalism applies, along with its associated higher geometry discussed above in the context of topological sigma models. Clearly, this is more general and goes beyond topological sigma models, pertaining any type of interacting gauge theory.  
Comprehensive treatments of higher gauge theory with both physical and mathematical perspectives include \cite{Baez:2005qu,Baez:2010ya,Grutzmann:2014hkn}. 

Before discussing some further elements of higher gauge theory, it is useful to comment right away on the relation to higher global symmetries, a subject which developed rather independently. Recall that the textbook treatment of gauge theory sets off with some global symmetry with a rigid transformation parameter. Due to Noether's theorem there are associated conservation laws and in particular conserved 1-form currents and charges. These currents can be coupled to the action of the theory with suitable 1-form background fields, essentially realizing the Noether procedure. Subsequently, and as long as no potential anomalies are present,{\footnote{We mainly refer here to 't Hooft anomalies, which can obstruct the gauging of a symmetry, see e.g. \cite{Cordova:2018cvg} for a recent review. The same reasoning also applies of course to ABJ anomalies.}} the global symmetry can be gauged by promoting the background fields to dynamical ones with their  kinetic term. 

The above standard picture of gauging a global symmetry has a higher analog. The starting point is the existence of higher global symmetries whose conserved currents are differential forms of degree greater than 1, see \cite{Gaiotto:2014kfa} and \cite{Sharpe:2015mja} for a review with a more complete set of references on the subject. As before, such symmetries can be gauged, first by coupling the currents to higher form background fields and then promoting them to dynamical ones. Thus although higher gauge theories have an independent history, one could also think of them as the gauged version of theories with higher global symmetry, as in the ordinary case. This is a topic with intensive activity in recent years.

Returning to higher gauge theory, let us discuss in some more detail the approach closest to the spirit of this review, in other words let us focus again on Q-manifolds and higher $E$-geometry. A complete treatment of this higher geometric viewpoint on higher gauge theory was given in \cite{Grutzmann:2014hkn}. There it is shown very generally that every higher gauge theory is in correspondence with a Q-manifold. This is of course in accord with what we discussed in the previous subsection, but it also holds for more general classes of gauge theories. On the other hand, this geometrical picture points to the direction of viewing every gauge theory essentially as a generalized sigma model of maps from a source space(time) to a suitable target space of fields. 

To be more specific, one can consider a tower of fields of any differential form degree, say scalar fields $X^{\mu}$, 1-forms $A^{a}$, 2-forms $B^{I}$, $\dots$, up to $p$-forms, with indices $\mu, a, I, \dots$ taking values from independent sets. Let us moreover collectively denote all fields as $\Phi^{\alpha}$, with $\alpha=\{\mu,a,I,\dots\}$. Then the field strength in this collective notation, comprising a variety of field strengths for each of the component fields, may be written as 
\begin{equation}
	F^{\alpha}=\mathrm{d} \Phi^{\alpha} + \sum_{r=1}^{p+1} \frac{1}{r!}f^{\alpha}{}_{\beta_1\dots \beta_r}(X)  \Phi^{\beta_1}\dots \Phi^{\beta_{r}}\,,
\end{equation}
where wedge products are understood wherever needed. In this expression, $f$'s are general field-dependent
structure functions. Consistency of the theory imposes additional constraints to them, typically in the form of Bianchi identities. For Yang-Mills theory, these would be simply the structure constants $f^{a}{}_{bc}$ of the gauge algebra, subject to the Jacobi identity. More generally, one should have the condition 
\begin{equation}
	\mathrm{d}F^{\alpha}\|_{F^{\alpha}=0}=0\,.
\end{equation}
Then \cite{Grutzmann:2014hkn} show that this condition, and its ``decomposition'' in a large number of complicated consistency conditions for the structure functions, is equivalent to the simple condition $Q^2=0$, where $Q$ is the degree-1 vector field defined on a suitable graded manifold ${\cal M}$ with coordinates $\xi^{\alpha}$. The specific form of $Q$ is 
\begin{equation}
	Q=\sum_{r=1}^{p+1}\frac 1{r!}f^{\alpha}{}_{\beta_1\dots\beta_r}(X)\xi^{\beta_1}\xi^{\beta_2}\dots\xi^{\beta_{r}}\frac{\partial}{\partial\xi^{\alpha}}\,.
\end{equation}
For instance, when ${\cal M}=E[1]$ with coordinates $\xi^{\alpha}=\{x^{\mu},\xi^{a}\}$ this results in the homological vector field \eqref{QLA} in Vaintrob's identification of Lie algebroids with Q-manifolds, provided that we identify the only nonvanishing structure functions as  $f^{\mu}{}_{a}=\rho^{\mu}{}_{a}$ and $f^{a}{}_{bc}=-C^{a}{}_{bc}$. 

Two remarks are in order here. First, the spirit of sigma models is evident from the fact that for the field content $\Phi^{\alpha}$ of differential form degrees $0, 1, \dots, p$ we introduced a graded manifold with coordinates $\xi^{\alpha}$ of degree $0, 1, \dots, p$.  As in sigma models where the scalar fields are viewed as pull-back functions from the coordinates of a target manifold via the sigma model map $X:\Sigma\to M$, namely $X^{\mu}=X^{\ast}(x^{\mu})$ where $x^{\mu}$ coordinates on $M$, we can think of the ``big map'' $\Phi:\hat{\Sigma}\to {\cal M}$ and 
\begin{equation}
	\Phi^{\alpha}=\Phi^{\ast}(\xi^{\alpha})\,,
\end{equation}
where $\hat{\Sigma}$ is a graded extension of $\Sigma$ as discussed previously, e.g. $T[1]\Sigma$. Note that both $\hat{\Sigma}$ and ${\cal M}$ are graded manifolds, and denoting their homological vector fields as $Q_{\hat{\Sigma}}$ and $Q_{\cal M}$, we can see that the field strengths are given as   
\begin{equation}
	F^{\alpha}={\cal F}(\xi^{\alpha})\,,\quad \text{where} \quad    {\cal F}=Q_{\hat{\Sigma}}\circ \Phi^{\ast}-\Phi^{\ast}\circ Q_{\cal M}\,.
\end{equation}
We will return to this generalized sigma model perspective in the next subsection in the context of tensor gauge theories. The second remark is that the above setting carries a strong $L_{\infty}$ flavour. $L_{\infty}$ algebras are generalizations of Lie algebras where, roughly, the binary Lie bracket and the Jacobi identity are replaced by a series of $n$-ary brackets and higher Jacobi identities, see \cite{Hohm:2017pnh} for an excellent introduction to their relation to field theory, \cite{Grutzmann:2014hkn} for their relation to the present context of higher gauge theory and \cite{Grewcoe:2020hyo} for an application in the context of Section \ref{sec2}.    

\subsection{Tensor Gauge Theories \& Universality of Form}\label{sec33} 

In the two physical applications we discussed so far, the spacetime (or world volume in the sigma model perspective) was the graded manifold $T[1]\Sigma$ of an ordinary manifold $\Sigma$. Recalling from Section \ref{sec2} that functions on this graded manifold are in correspondence with differential forms, this is expected in view of the fact that we have been dealing with higher form fields and their gauge theory. However, in physics we do not encounter only scalar fields and differential forms, but also higher tensor fields. The obvious one is the graviton, a symmetric tensor field. Nevertheless, higher spin fields, higher symmetric tensors for example, can also be relevant in several contexts and moreover they can often arise through generalizations of electric/magnetic duality. For example, the so-called Curtright field, a 3-index tensor with $(2,1)$ mixed symmetry (antisymmetric in two indices but not in all three) arises as the dual of the graviton in 5D \cite{Hull:2001iu}. Gauge theories for this field and $(p,q)$ generalizations thereof were studied  in \cite{Curtright:1980yk}. From a different physical motivation, tensor gauge theories find applications in condensed matter systems with subsystem symmetries, such as fractons, see e.g. the review \cite{Pretko:2020cko} for pointers in this direction.   

In view of the above and in relation to the rest of this review, one could naturally ask what is the $E$-geometry or Q-manifold description for theories with mixed symmetry tensor fields in their field content. Although a complete mathematical description is still lacking, some steps in this direction were taken in \cite{Chatzistavrakidis:2019len,Chatzistavrakidis:2020kpx,Chatzistavrakidis:2021dqg}. In this approach, one introduces a graded manifold with two differentials $Q_1$ and $Q_2$, both being homological and in addition commuting, 
\begin{equation}
	(Q_1)^2=0=(Q_2)^2 \qquad \text{and}\qquad [Q_1,Q_2]=0\,.
\end{equation}
In the simplest possible setting, one can then model the spacetime manifold as a ``double copy''{\footnote{One should not take this as having literally any relation to the matching of scattering amplitudes between gravity and Yang-Mills.}} $T[1,0]\Sigma\oplus T[0,1]\Sigma$, indicating that the coordinates on this manifold are two sets of fermionic ones $\theta_1^{\mu}$ and $\theta_2^{\mu}$ aside the bosonic ones $x^{\mu}$. In the same way as functions on $T[1]\Sigma$ correspond to differential forms, functions on this extended graded manifold correspond to bipartite mixed symmetry tensors of type $(p,q)$. Of course, this picture includes scalar fields and differential forms already and in addition it can be extended to higher spin fields once more femionic coordinates are introduced.{\footnote{It is amusing to note that the more symmetric slots one adds to a tensor, the more antisymmetric coordinates are introduced to describe them in terms of Q-manifolds. I thank Peter Schupp for emphasizing this.}}  

With the above simple starting point, one can build a graded differential calculus on the extended source space and develop an extended $(Q_1,Q_2)$-manifold perspective to higher tensor gauge theory. 
In view of the length constraints of this mini-review and skipping details that may be found in the original papers, let us briefly mention the key points of this approach. The main advantage is that it leads to a ``universality of form'' for gauge theories with different field content and spins. Restricting here to bipartite mixed symmetry tensors, this means that there exists a universal form of the action of a linear or algebraically nonlinear{\footnote{This means that Lagrangians whose nonlinearity is due to being an algebraic functional of the free kinetic and theta terms are included, but Yang-Mills type theories or full general relativity have not been fully described in this form yet.}} (gauge, in the massless case) theory that is the same for scalar fields, differential forms and $(p,q)$ tensors and it reads
\begin{equation}
	S_{\text{uni}}[\omega]=\int\left(\frac 12 G_{\mu\nu}(\omega,\mathrm{d}\omega)\mathrm{d}\omega^{\mu}\star\mathrm{d}\omega^{\nu}+\frac 12B_{\mu\nu}(\omega,\mathrm{d}\omega)\mathrm{d}\omega^{\mu}\mathrm{d}\omega^{\nu}\right)+S_{\text{m}}+S_{\text{int}}\,,
\end{equation}
for a suitable $\star$ operator that defines a good inner product \cite{Chatzistavrakidis:2019len} and for an integral that refers to spacetime as well as to the Berezin integration over the fermionic coordinates. In this universal action, we have included a universal mass term given simply as 
\begin{equation}\label{Suni}
	S_{\text{m}}\propto \int \omega\star\omega\,,
\end{equation}
and an interaction term that can include for instance higher derivative Galileon-type interactions for tensor fields as found in \cite{Chatzistavrakidis:2016dnj}. In the multifield case one includes background fields $G$ (symmetric 2-tensor) and $B$ (symmetric in $4n$ and antisymmetric in $4n+2$ dimensions). Focusing only on the kinetic (or also the mass) term and climbing up the ladder of options, this action is precisely the massless (or massive) scalar or the Maxwell (Proca) theory or the linearized Einstein-Hilbert (or massive Fierz-Pauli) or the Curtright (massive Curtright) theory, etc., all included in the same starting point.     

Notably and in relation to the discussion in Section \ref{sec32}, such actions and their first-order analogs were used to identify higher global symmetries and their 't Hooft anomalies in a universal way and also to offer a geometric interpretation to higher global symmetries as generalized isometries along graded vector fields on a graded target space or its 1st jet space thereof \cite{Chatzistavrakidis:2021dqg}. This perspective is along the lines of thinking of such theories universally as generalized sigma models, also in the case of higher spins such as the graviton. This way we see once more the close interplay between Q-manifolds and field theory, this time for higher tensor fields.


\paragraph{Acknowledgements.}
Work supported by the Croatian Science Foundation Project ``New Geometries for Gravity and Spacetime'' (IP-2018-01-7615).

\end{document}